\documentclass[preprint2]{aastex}

\usepackage{amsmath,alltt}
\newcommand{\grad}{\nabla}
\newcommand{\p}{\partial}
\newcommand{\pt}{\partial t}

\newcommand{\px}{\partial x}
\newcommand{\te}{\tilde{e}}
\newcommand{\tv}{\tilde{v}}
\newcommand{\De}{\Delta e}
\newcommand{\Dt}{\Delta t}
\newcommand{\Dtau}{\Delta\tau}
\newcommand{\Dx}{\Delta x}
\newcommand{\Dv}{\Delta v}
\newcommand{\Dtv}{\Delta\tv}
\newcommand{\ba}{\boldsymbol{a}}
\newcommand{\bk}{\boldsymbol{k}}
\newcommand{\bu}{\boldsymbol{u}}
\newcommand{\bvel}{\boldsymbol{v}}
\newcommand{\bx}{\boldsymbol{x}}
\newcommand{\bDv}{\boldsymbol{\Dv}}
\newcommand{\btv}{\boldsymbol{\tv}}
\newcommand{\bL}{\boldsymbol{L}}
\newcommand{\bLa}{\bL^a}
\newcommand{\bLe}{\bL^e}
\newcommand{\bLg}{\bL^g}
\newcommand{\bLv}{\bL^v}
\newcommand{\bF}{\boldsymbol{F}}
\newcommand{\bFa}{\bF^a}
\newcommand{\bFe}{\bF^e}
\newcommand{\bS}{\boldsymbol{S}}
\newcommand{\bSg}{\bS^g}
\newcommand{\bSv}{\bS^v}

\slugcomment{}

\shorttitle{Moving Frame Hydrodynamic Algorithm}
\shortauthors{Trac \& Pen}

\begin{document}


\title{A MOVING FRAME ALGORITHM FOR HIGH MACH NUMBER HYDRODYNAMICS}


\author{Hy Trac}
\affil{Department of Astronomy and Astrophysics, University of Toronto, Toronto, ON M5S 3H8, Canada}
\email{trac@cita.utoronto.ca}

\and

\author{Ue-Li Pen}
\affil{Canadian Institute for Theoretical Astrophysics, 60 St. George Street, Toronto, ON M5S 3H8, Canada}
\email{pen@cita.utoronto.ca}



\begin{abstract}
We present a new approach to Eulerian computational fluid dynamics that is designed to work at high Mach numbers encountered in astrophysical hydrodynamic simulations.  The Eulerian fluid conservation equations are solved in an adaptive frame moving with the fluid where Mach numbers are minimized.  The moving frame approach uses a velocity decomposition technique to define local kinetic variables while storing the bulk kinetic components in a smoothed background velocity field that is associated with the grid velocity.  Gravitationally induced accelerations are added to the grid, thereby minimizing the spurious heating problem encountered in cold gas flows.  Separately tracking local and bulk flow components allows thermodynamic variables to be accurately calculated in both subsonic and supersonic regions.  A main feature of the algorithm, that is not possible in previous Eulerian implementations, is the ability to resolve shocks and prevent spurious heating where both the preshock and postshock Mach numbers are high.  The hybrid algorithm combines the high resolution shock capturing ability of the second-order accurate Eulerian TVD scheme with a low-diffusion Lagrangian advection scheme.  We have implemented a cosmological code where the hydrodynamic evolution of the baryons is captured using the moving frame algorithm while the gravitational evolution of the collisionless dark matter is tracked using a particle-mesh N-body algorithm.  The cosmological {\it MACH} code is highly suited for simulating the evolution of the IGM where accurate thermodynamic evolution is needed for studies of the Lyman alpha forest, the Sunyaev-Zeldovich effect, and the X-ray background.  Hydrodynamic and cosmological tests are described and results presented.  The current code is fast, memory-friendly, and parallelized for shared-memory machines.
\end{abstract}




\section{Introduction}

Computational fluid dynamics (CFD) is a powerful approach to simulating the complex fluid flow present in astrophysical hydrodynamics.    Modeling astrophysical structure formation and the dynamics of astrophysical systems is a challenging problem because the nonlinear gasdynamical processes can span a large range in scale, mass, and energy.  Furthermore, strong shocks, gravitational collapse, and radiative feedback can occur and play an important role in the evolution of the gas.   In astrophysics, CFD has also been coupled to N-body dynamics and to magnetohydrodynamics (MHD).  In particular, combined hydro plus N-body simulations have been successfully applied in cosmology and galactic dynamics to simultaneously probe the interaction between the baryonic gas and collisionless dark matter.  Continuing advancement in the field of CFD is making it tractable to robustly probe the nonlinear physics involved.  Both Eulerian and Lagrangian methods have been developed with emphasis on high-resolution capturing of shocks, prevention of numerical instabilities, and having high dynamic range. 

In this paper we focus on Eulerian CFD, where the standard approach is to discretize time into discrete steps and space into finite volumes or cells.  In the simplest case, the integral Euler equations are solved on a Cartesian lattice by computing the flux of conserved quantities like mass, momentum, and energy across grid cell boundaries.  Traditionally, Eulerian CFD is recognized for its superior shock-capturing ability.  In particular, the flux-conservative scheme based on the total variation diminishing (TVD) condition \citep{harten83}  has been demonstrated to provide high resolution capturing of shocks while preventing unphysical oscillations \citep[See][for a review]{tp03}.  The Eulerian approach has a large dynamic range in mass but not in length, opposite to that of Lagrangian schemes.  The large dynamic range in mass at all scales allows both large and small scales to be probed simultaneously.  In addition, Eulerian algorithms are computationally very fast and memory-friendly, allowing one to optimally use computing resources.  Some examples of successful TVD codes are \citet{ryu93}, \citet{gnedin95}, and \citet{pen98}.  

One of the main challenges in simulating complex fluid flow is the capturing of shocks in the presence of high Mach numbers.  For example in cosmological simulations of the intergalactic medium (IGM), there are large-scale velocity fields on the order of 1000 km/s and the typical sound speed in these bulk flows is $\sim10$ km/s.  At Mach numbers $\sim100$, the ratio of the thermal energy to the kinetic energy is $\sim10^{-4}$.  In Eulerian CFD, the standard practice is to calculate the thermal energy by subtracting the kinetic energy from the total energy, but this calculation can be inaccurate, especially near shocks.  The physical solutions for shock waves are discontinuous, making approximations near the shock fronts only first-order accurate.  Thus, the cancellation error of subtracting the kinetic energy from the total energy will typically result in significant errors.  The poor tracking of the thermal energy in supersonic regions is known as the high Mach number problem in Eulerian CFD. 

\citet{ryu93} and \citet{bryan95} have attempted to overcome the high Mach number problem by choosing to solve an alternative equation rather than the total energy conservation equation near supersonic regions in order to calculate the thermal energy more accurately.  The former solves an entropy equation while the latter directly solves the internal energy equation.  Both methods appear to work when subjected to standard hydro tests, but these tests are static ones in which the medium is initially at rest.  The static tests are idealized in that they involve strong shocks where the postshock fluid is subsonic.  But in practice, regions experiencing shocks are often dynamical and can be embedded in fast-moving flows.  For cosmological applications such as the study of the Lyman alpha forest, we often need to resolve modest shocks with pairwise velocities on the order of $\sim 10$ km/s in a large-scale flow moving $\sim 10-100$ times faster.  Capturing the thermodynamic evolution is very difficult when the postshock Mach numbers also remain high.  For such shocks, a total energy conservation equation must be solved to achieve the correct shock jump conditions.  But \citet{ryu93} and \citet{bryan95} do not use the total energy equation for thermodynamic calculations since the postshock Mach numbers are also high.  The cosmological velocity field is dominated by large-scale flows spanning $\sim10^2$ Mpc and large simulation box sizes are needed to accurately capture the evolution of large-scale structure in the IGM.  This puts more demand on the Mach number dynamic range of cosmological hydrodynamic codes.

We present a new approach to doing Eulerian CFD that is designed to work at high Mach numbers.  The {\it MACH} code uses a hybrid Eulerian-Lagrangian algorithm that solves the Eulerian fluid equations in an adaptive frame moving with the fluid, allowing local quantities to be directly tracked.  
A unique feature of the algorithm is to be able to resolve shocks and prevent spurious heating where both the preshock and postshock Mach numbers are high.  In $\S$\ref{sec:eulerhydro} we review the formalism for Eulerian hydrodynamics and present a new approach to solving the high Mach number problem.  In $\S$\ref{sec:mach} we describe the new algorithm for Eulerian CFD at high Mach numbers.  An in-depth pedagogical review of Eulerian CFD is presented in \citet{tp03} and we encourage the reader to use this source for supplementary details.  Hydrodynamic tests are described in $\S$\ref{sec:hydrotests}.  We have modified standard, static hydro tests to be dynamic in order to study the ability of the moving frame algorithm to capture shocks in fast-moving flows.  In $\S$\ref{sec:cosmo}, we describe an adaptation of the {\it MACH} code for cosmological applications.  The hydrodynamic evolution of the baryons is handled using the {\it MACH} algorithm while the gravitational evolution of the collisionless dark matter is tracked using a particle-mesh (PM) N-body algorithm.  Cosmological test are described and results presented.

\section{Eulerian Hydrodynamics}
\label{sec:eulerhydro}

In this section, we present the formalism for simulating astrophysical fluids.  The Euler equations are a system of conservation laws which govern hydrodynamics.  In differential conservation form, the continuity equation, momentum equation, and energy equation are given as
\begin{align}
\label{eqn:E1}
\frac{\p\rho}{\pt}+\frac{\p}{\px_j}(\rho v_j)&=0,\\[8pt]
\label{eqn:E2}
\frac{\p(\rho v_i)}{\pt}+\frac{\p}{\px_j}(\rho v_iv_j+P\delta_{ij})&=-\rho\frac{\p\phi}{\px_i},\\[8pt]
\label{eqn:E3}
\frac{\p e}{\pt}+\frac{\p}{\px_j}[(e+P)v_j]&=-\rho v_i\frac{\p\phi}{\px_i}.
\end{align}
where the physical state of the fluid is specified by its density $\rho$, velocity field $\bvel$, and total energy density
\begin{equation}
e=\frac{1}{2}\rho v^2+\epsilon.
\label{eqn:totalenergy}
\end{equation}
In standard practice, the thermal energy $\epsilon$ is evaluated by subtracting the kinetic energy from the total energy.  For an ideal gas, the pressure $P$ is related to the thermal energy by the equation of state
\begin{equation}
P=(\gamma-1)\epsilon,
\end{equation}
where $\gamma$ is the ratio of specific heats.  Poisson's equation
\begin{equation}
\grad^2\phi=4\pi G\rho,
\label{eqn:poisson}
\end{equation}
relates the gravitational potential $\phi$ to the density field.

\subsection{Frame Decomposition}
\label{sec:frame}

In the Eulerian approach, the fluid variables are defined with respect to the static grid frame of the simulation box.  The high Mach number problem arises when the bulk flow components in the velocity field and total energy outweigh the corresponding local components.  In the Lagrangian approach, the fluid equations are solved in the frame of the fluid and local variables can be naturally computed.  We consider a hybrid scheme in which the Eulerian conservation equations are solved in a local frame moving with the flow and local quantities can be directly tracked.  Our approach will explicitly decompose the equations into a smooth flow, whose solution can be obtained by any simple finite-difference scheme, and a nonlinear component at low Mach numbers that can be solved by a traditional flux-conservative TVD solver.  The TVD flux-conservative schemes are critical to capture shocks in the presence of discontinuities, where derivatives are ill-defined.  This approach has the advantage of being able to effectively capture shocks like in standard grid-based schemes, but does not suffer from the high Mach number problem.

We can choose a frame that is locally moving with the fluid using a velocity decomposition technique.  The velocity field of the fluid,
\begin{equation}
\bvel=\bDv+\btv,
\label{eqn:globalvel}
\end{equation}
is decomposed into a local term $\bDv(\bx)$ and a smoothed background term $\btv(\bx)$.  The local velocity is a peculiar velocity with respect to the Eulerian grid and the smoothed background velocity is associated with the velocity of the grid cells.  This construction results in a moving background reference frame that can approximate the bulk flow.  The total energy density,
\begin{equation}
e=\De+\te,
\end{equation}
can also be separated into a local term $\De(\bx)$ and a grid term $\te(\bx)$.  The local total energy density,
\begin{equation}
\De(\bx)=\frac{1}{2}\rho\Dv^2+\varepsilon,
\label{eqn:localenergy}
\end{equation}
is defined as the sum of the local kinetic energy and the thermal energy, analogous to equation (\ref{eqn:totalenergy}).  The grid energy density,
\begin{equation}
\te=\frac{1}{2}\rho\tv^2+\rho\bDv\cdot\btv,
\label{eqn:gridenergy}
\end{equation}
is uniquely defined in terms of the other fluid variables and is not a free variable.  This decomposition allows us to individually keep track of local and bulk terms.  The thermal energy can be determined more accurately since there are no bulk flow components in equation (\ref{eqn:localenergy}).

In the standard Euler equations (eq. [\ref{eqn:E1}-\ref{eqn:E3}]), the free fluid variables are the density $\rho$, velocity $\bvel$, and total energy density $e$.  With the introduction of equations (\ref{eqn:globalvel}) to (\ref{eqn:gridenergy}), the state of the fluid is now described by the density $\rho$, local velocity $\bDv$, grid velocity $\btv$, and local total energy density $\De$.  The addition of the grid velocity requires an additional equation,
\begin{equation}
\frac{\p\tv_i}{\pt}+\Dv_j\frac{\p\tv_i}{\px_j}+\tv_j\frac{\p\tv_i}{\px_j}=0,
\label{eqn:gridvel}
\end{equation}
to govern its time evolution.  This advection equation for the grid velocity is similar to Burger's equation.  In this frame decomposition, the Euler equations are then given by
\begin{multline}
\label{eqn:eE1}
\frac{\p\rho}{\pt}+\frac{\p}{\px_j}(\rho\Dv_j)+\frac{\p}{\px_j}(\rho\tv_j)=0,\\
\end{multline}
\begin{multline}
\label{eqn:eE2}
\frac{\p(\rho\Dv_i)}{\pt}+\frac{\p}{\px_j}(\rho\Dv_i\Dv_j+P\delta_{ij})+\frac{\p}{\px_j}(\rho\Dv_i\tv_j)\\
=-\rho\frac{\p\phi}{\px_i},
\end{multline}
\begin{multline}
\label{eqn:eE3}
\frac{\p\De}{\pt}+\frac{\p}{\px_j}[(\De+P)\Dv_j]+\frac{\p}{\px_j}(\De\tv_j)\\
=-\rho (\Dv_i+\tv_i)\frac{\p\phi}{\px_i}-P\frac{\p\tv_j}{\px_j},
\end{multline}
%\begin{align}
%\label{eqn:eE1}
%\frac{\p\rho}{\pt}+\frac{\p}{\px_j}(\rho\Dv_j)+\frac{\p}{\px_j}(\rho\tv_j)&=0,\\[8pt]
%\label{eqn:eE2}
%\frac{\p(\rho\Dv_i)}{\pt}+\frac{\p}{\px_j}(\rho\Dv_i\Dv_j+P\delta_{ij})+\frac{\p}{\px_j}(\rho\Dv_i\tv_j)&=-\rho\frac{\p\phi}{\px_i},\\[8pt]
%\label{eqn:eE3}
%\frac{\p\De}{\pt}+\frac{\p}{\px_j}[(\De+P)\Dv_j]+\frac{\p}{\px_j}(\De\tv_j)&=-\rho(\Dv_i+\tv_i)\frac{\p\phi}{\px_i}-P\frac{\p\tv_j}{\px_j},
%\end{align}
and completed by equation (\ref{eqn:gridvel}).  The velocity decomposition is reflected in the structure of the expanded Euler equations (eq. [\ref{eqn:gridvel}-\ref{eqn:eE3}]).  The first gradient term on the left hand side of each of the equations describes the flux of the free variables in the local frame where the fluid moves at the local velocity $\bDv$.  We will refer to these terms as the Euler terms.  The second gradient term describes the advection of the same quantities as the underlying background frame moves at the grid velocity $\btv$.  We call these terms the advection terms.  The expanded Euler equations reduce to the standard ones when there is no background velocity field.  In the case where the background velocity is a constant, the expanded Euler equations are just the standard ones with a simple Galilean transformation.  For a nonuniform background velocity field, the energy equation (eq. [\ref{eqn:eE3}]) has an additional source term because the compression of the grid introduces a Coriolis term.

\section{MACH Algorithm}
\label{sec:mach}

We now describe a hybrid algorithm for solving the expanded Euler equations.  To simplify the notation, we write equations (\ref{eqn:eE1}) to (\ref{eqn:eE3}) in vector form:
\begin{equation}
\frac{\p\bu}{\pt}+\frac{\p\bFe_j}{\px_j}+\frac{\p\bFa_j}{\px_j}=\bSg_j+\bSv_j,
\label{eqn:veE}
\end{equation}
where $\bu=(\rho,\rho\bDv,\De)$, $\bF$ represents flux terms, and $\bS$ represents source terms.  The flux terms in the Euler and advection operations are represented by $\bFe$ and $\bFa$, respectively.  Gravitational source terms are included in $\bSg$ while source terms arising from the velocity decomposition are  found in $\bSv$.  The expanded Euler equations are conveniently solved using an operator-splitting technique \citep{strang68}.  Let the operator $\bL$ represent the updating of $\bu(\bx,t)$ to $\bu(\bx,t+\Dt)$ for a given operation.  We first do a forward sweep:
\begin{equation}
\bu^{t+\Dt}=\bLg\bLa\bLv\bLe\bu^t,
\end{equation} 
and then perform a reverse sweep:
\begin{equation}
\bu^{t+2\Dt}=\bLe\bLv\bLa\bLg\bu^{t+\Dt},
\end{equation}
using the same time step $\Dt$ to obtain second-order accuracy.  The operator-splitting technique is also useful for solving multi-dimensional problems.  In three dimensions, the Euler equations can be dimensionally split into three separate one-dimensional equations that can be solved sequentially.  For the forward and reverse sweeps, we have
\begin{align}
\label{eqn:ds1}
\bu^{t+\Dt}&=\bLg(\bLa\bLv\bLe)_z(\bLa\bLv\bLe)_y(\bLa\bLv\bLe)_x\bu^t,\\
\label{eqn:ds2}
\bu^{t+2\Dt}&=(\bLe\bLv\bLa)_x(\bLe\bLv\bLa)_y(\bLe\bLv\bLa)_z\bLg\bu^{t+\Dt},
\end{align}
and this splitting scheme is second-order accurate.  The gravity operator is not dimensionally split.

\subsection{Computational Fluid Dynamics}
\label{sec:cfd}

The dimensional-splitting scheme effectively renders the problem of solving the multi-dimensional Euler equations into a one-dimensional one.  For ease of illustration, we present solutions for a one-dimensional problem.  Generalization to higher dimensions is straightforward following the prescription given by equations (\ref{eqn:ds1}) and (\ref{eqn:ds2}).

The standard approach to Eulerian CFD is to discretize time into discrete steps and space into finite volumes or cells, where the cell-averaged values of hydrodynamic variables are stored.  In practice, the quantity $\bu_n^t\equiv\bu(x_n,t)$ and fluxes $\bF_n^t\equiv\bF(x_n,t)$ are defined at integer grid cell centers $x_n$.  The challenge is to use the cell-averaged values to determine the fluxes at cell boundaries.

\subsubsection{The Euler operation}

The Euler operation involves computing the flux of the free variables in the local frame where the fluid moves at the local velocity $\Dv$.  Here we solve the system of equations:
\begin{align}
\label{eqn:eulerop1}
\frac{\p\bu}{\pt}+\frac{\p\bFe}{\px}&=0,\\[8pt]
\label{eqn:eulerop2}
\frac{\p\tv}{\pt}+\Dv\frac{\p\tv}{\px}&=0.
\end{align}
Equation (\ref{eqn:eulerop1}) is a vector conservation law that can be robustly solved using a second-order accurate TVD flux-assignment scheme with a second-order Runge-Kutta time integration technique \citep[See][]{tp03}.  The general solution has the flux-conservative form
\begin{equation}
\bu_n^{t+\Dt}=\bu_n^t-\left(\frac{\bF_{n+1/2}^{t+\Dt/2}-\bF_{n-1/2}^{t+\Dt/2}}{\Dx}\right)\Dt,
\end{equation}
where the half-step fluxes $\bF^{t+\Dt/2}$ at cell boundaries are determined using the TVD flux assignment scheme.  The procedure for assigning fluxes at cell boundaries based on cell-centered fluxes is difficult for the Euler operation.  Due to the inclusion of gas pressure in the fluxes, the direction of flow depends on both the local velocity $\Dv$ and the local sound speed $c_s$ and is not straightforward to determine.  We implement the relaxing TVD scheme \citep{jx95} with a van Leer flux limiter \citep{vl74} to provide high-resolution capturing of shocks.  The relaxing scheme offers an accurate and robust procedure for decomposing the flow into left and right moving waves, thereby making flux assignment straightforward \citep[See][]{tp03}.  The relaxing TVD scheme has been successfully implemented for simulating astrophysical fluids by \citet{pen98} and \citet{tp03}.  

Equation (\ref{eqn:eulerop2}) is coupled to equation (\ref{eqn:eulerop1}) via the local velocity $\Dv$.  We solve the advection of the grid velocity using a second-order Runge-Kutta scheme:
\begin{equation}
\tv_n^{t+\Dt}=\tv_n^t-\Dv_n^{t+\Dt/2}\left(\frac{\tv_{n+1}^{t+\Dt/2}-\tv_{n-1}^{t+\Dt/2}}{2\Dx}\right)\Dt,
\label{eqn:eulerop3}
\end{equation}
where the half-step local velocity $\Dv^{t+\Dt/2}$ comes from the half-step value in the relaxing TVD solution.  By construction, the grid velocity is smooth and a simple finite difference can be used to approximate the gradient.  When fully expanded, equation (\ref{eqn:eulerop3}) is similar to the second-order Lax-Wendroff equation \citep{lw60}.  We choose the Runge-Kutta scheme rather than the Lax-Wendroff scheme because the former provides an effective way to simultaneously solve equations (\ref{eqn:eulerop1}) and (\ref{eqn:eulerop2}).

In the energy equation (eq. [\ref{eqn:eE3}]), the source term arising from the velocity decomposition is associated with the Euler operation and not the advection operation.  Therefore, we calculate the solution for the source term using a second-order Runge-Kutta scheme where the half-step pressure $P^{t+\Dt/2}$ and half-step grid velocity $\tv^{t+\Dt/2}$ comes from the half-step values in the solutions for equations (\ref{eqn:eulerop1}) and (\ref{eqn:eulerop2}).  This additional Coriolis flux is from the compression of the grid.  Since the grid velocity is smooth, the contributions from the source term will be small.  

The total variation diminishing condition \citep{harten83} is a nonlinear stability condition.  The relaxing scheme for the Euler operation is TVD and stable provided that Courant or CFL number satisfies
\begin{equation}
\lambda\equiv\frac{{\rm max}(c_n)\Dt}{\Dx}<1,
\end{equation}
where $c\equiv|\Dv|+c_s$ is called the freezing speed and $c_s$ is the sound speed.  The freezing speed is constructed to be greater than the largest eigenvalue of the flux Jacobian $\p\bFe/\p\bu$.  Depending on the application, the CFL number is normally chosen in the range $0.5<\lambda<0.9$.

\subsubsection{The advection operation}

The advection operation involves transporting the free variables as the underlying background frame moves at the grid velocity $\tv$.  The advection system of equations:
\begin{align}
\label{eqn:advectop1}
\frac{\p\bu}{\pt}+\frac{\p\bFa}{\px}&=0,\\[8pt]
\label{eqn:advectop2}
\frac{\p\tv}{\pt}+\tv\frac{\p\tv}{\px}&=0,
\end{align}
can be solved using the same method implemented for the Euler operation, but advection is more accurately solved using a Lagrangian approach to minimize numerical diffusion.  Note that equation (\ref{eqn:advectop2}) has the form of Burger's equation, which describes the self-transport of a velocity field.  In the Lagrangian step, the grid cells or zones are advected along lines of constant mass, where zone centres satisfy the equation
\begin{equation}
x_n^{t+\Dt}=x_n^t+\tv_n^t\Dt.
\label{eqn:zones}
\end{equation}
By construction, the starting distribution of Eulerian grid cells is regular, with equal volume cells.  In general, the resulting distribution of Lagrangian zones will be irregular, with unequal volume zones.  The Lagrangian zones are then remapped back onto the Eulerian grid for the next set of hydrodynamic operations.  

Note that for a specific advection velocity in equation (\ref{eqn:zones}), zone centres can be transported to locations that coincide with grid cell centres, making the remap process straightforward.  This observation is the central point of how the remap technique works.  We can choose the advection velocity for the Lagrangian step to be
\begin{equation}
b_n^t\equiv\tv_n^t-a_n^t=\frac{{\rm Round}(\tv_n^t\Dt)}{\Dt},
\end{equation}
where the residual term
\begin{equation}
a_n^t=\tv_n^t-\frac{{\rm Round}(\tv_n^t\Dt)}{\Dt}\,
\end{equation}
is chosen such that the distance a zone is transported is an integer multiple of the regular grid cell separation $\Dx=1$.  First, we solve the advection system of equations (eq. [\ref{eqn:advectop1}-\ref{eqn:advectop2}]), but for an advection velocity equal to $a$ rather than $\tv$.  This is done on the Eulerian grid using the Eulerian approach.  Second, the Lagrangian step is carried out by transporting the quantities to their new location using the advection velocity $b$.

The first step of the remap process can be solved using the Eulerian scheme devised for the Euler operation.  However, we make one modification to speed up the algorithm.  Since equation (\ref{eqn:advectop1}) is purely an advection equation, we can explicitly check the advection velocity to determine the direction of flow and assign fluxes accordingly.  The relaxing scheme is replaced with a standard upwind TVD scheme that is computationally less expensive \citep[See][]{tp03}.  Note that advection at the residual velocity $a$ automatically satisfies the CFL stability condition because for any given time step, the residual velocity is chosen such that the fluid is moved no more than half a grid cell.

In the second step, each Lagrangian zone is mapped to the corresponding Eulerian grid cell.  Multiple zones can be mapped to the same grid cell and as a result, some grid cells may be left empty.  Rather than having 2 zones be mapped to 2 cells with a gap inbetween, we re-adjust the velocities $a$ and $b$ so that the 2 zones get mapped to 3 cells, thereby avoiding having empty cells.  We impose the restriction that near divergent regions, the mapping process cannot result in the circumstance where an empty cell is adjacent to another empty cell.  This criterion is satisfied if
\begin{equation}
\Dt<\frac{1}{{\rm max}(|b_{n+1}-b_n|)},
\label{eqn:gridtimestep}
\end{equation}
Since the grid velocity $\tv$ is smooth and its gradient small, this restriction rarely enforces the limiting time step.  In almost all practical cases, the time step in the Euler operation will satisfy equation (\ref{eqn:gridtimestep}).

The two-step remap process is equivalent to advection at the velocity $\tv$, but this particular scheme offers some advantages over the alternative of solving the advection system using the Eulerian approach described for the Euler operation.  Since the grid velocity is constructed to approximate the bulk flow, it can be relatively large compared to the local velocity $\Dv$ or the local sound speed $c_s$ and would be a limiting factor in the length of the time step.  However, we have seen that the two-step remap process effectively has no limiting time step when the grid velocity is smooth.  Using longer time steps minimizes the numerical diffusion and shortens simulation run time.

\subsection{Frame change}

We call the procedure of constructing a new smoothed background velocity field a frame change.  Again, the dimensional-splitting scheme effectively renders this process into a one-dimensional one.  Since the flux sweep in the $x$ direction only involves the gradients $\p/\px$, the smoothing only has to be done in one direction at a time.  We first illustrate the solution for a one-dimensional problem and then generalize it to higher dimensions.

At the beginning of both the forward and reverse sweeps, we construct a smoothed background velocity field $\tv$ by smoothing the global velocity field $v$:
\begin{equation}
\tv_n=\frac{\sum_{m=1}^N v_m w_m s(x_n-x_m)}{\sum_{m=1}^N w_m s(x_n-x_m)},
\end{equation}
where $N$ is the total number of grid cells, $w(x)$ is a weight function, and $s(x)$ is a smoothing kernel.
We choose a Gaussian kernel
\begin{equation}
s(x)=\frac{1}{\sqrt{2\pi}R}\exp\left(-\frac{x^2}{2R^2}\right),
\end{equation}
with smoothing radius $R$ that will wash out the local peculiar velocity and leave the bulk velocity.  The discrete convolution is computed using one dimensional fast Fourier transforms (FFTs).  

The weight function can be set to unity in the simplest case, but it may be desirable to weight the contributions of the global velocity field not just on locality but also on the properties of the physical environment.  In general, cold regions are more likely to suffer from the high Mach number problem simply due to the fact that the sound speed is small.  Consider the case of a moving shock front.  If we choose the reference frame to be that of the shock, then the cold gas in front of the shock will be moving supersonically in this frame and have high Mach numbers.  However, if we choose the reference frame to be that of the cold ambient medium, then the ambient gas will be subsonic.  While the postshock medium will be moving fast with respect to this reference frame, it will have low Mach numbers since the postshock gas is hot.  Hence, we adopt a weight function
\begin{equation}
w_n=\frac{1}{\sqrt{{\rm max}(T_n,T_{min})}},
\label{eqn:smoothweight}
\end{equation}
where $T$ is the temperature distribution of the gas.  In practice, we impose a minimum temperature $T_{min}$ in equation (\ref{eqn:smoothweight}) to prevent inaccurate weighting due to the possibility of unphysical temperatures or spurious oscillations in the temperature distribution.

During a frame change, mass is automatically conserved since the definition of mass is absolute.  Momentum and energy conservation requires adjusting the local velocity $\Dv$ and local total energy density $\De$.  If the change in the grid velocity is $\Dtv_n$ for cell $n$, then we have
\begin{align}
\Dv_n^f&=\Dv_n^i-\Dtv_n,\\[8pt]
\De_n^f&=\De_n^i+\frac{1}{2}\rho_n(\Dtv_n)^2-\rho\Dv_n^i\Dtv_n.
\end{align}
where $i$ and $f$ label the initial and final states with respect to the frame change.

For multi-dimensional problems, all components of the global velocity $\bvel$ need to be decomposed, but the smoothing only has to be done in the direction of the flux sweep. For a three-dimensional problem, the splitting scheme given by equations (\ref{eqn:ds1}) and (\ref{eqn:ds2}) involves 6 flux sweeps, each of which is preceded by a frame change.

\subsection{Gravity}
\label{sec:gravity}

In the expanded Euler equations, the gravitational source terms are found on the right-hand side of the momentum and energy equations, just like for the standard Euler equations.  However, we can choose to remove the gravitational source terms in the momentum and energy equation and equivalently add a gravitational source term to the grid velocity equation,
\begin{equation}
\frac{\p\tv_i}{\pt}+\Dv_j\frac{\p\tv_i}{\px_j}+\tv_j\frac{\p\tv_i}{\px_j}=-\frac{\p\phi}{\px_i}.
\label{eqn:gravitygridvel}
\end{equation}
Gravity is a global process and it is more logical and advantageous to add the gravitationally induced changes to the grid rather than to the local quantities.  An important point made by \citep{ryu93} is that since the gravitational energy can be comparable to the kinetic energy and much larger than the thermal energy, numerical errors in calculating the gravitational effects can significantly cause spurious heating of cold gas.  By adding the gravitationally induced changes to the grid, we can minimize the spurious heating problem.  This is another example of how the velocity decomposition technique allows one to separate local and global components.

Poisson's equation (eq. [\ref{eqn:poisson}]) relates the gravitational potential to the density field and the general solution can be written as
\begin{equation}
\phi(\bx)=\int\rho(\bx')w(\bx-\bx')d^3x',
\label{eqn:phi}
\end{equation}
where the isotropic kernel is given by
\begin{equation}
w(r)=-\frac{G}{r}.
\end{equation}
In the discrete case, the integral in equation (\ref{eqn:phi}) becomes a sum and Poisson's equation can be solved using FFTs to do the convolution.  The force terms $f_i\equiv-\p\phi/\px_i$ on the right hand side of the Euler equations are calculated by finite differencing the potential.  The real space kernel $w(r)$ is constructed on the grid with the zero point satisfying
\begin{equation}
f(1)=-\frac{w(2)-w(0)}{2\Dx}=-G.
\end{equation}
By choosing $w(0)=-2.5G$, we make the force exact at a separation of one grid cell.  Alternatively, the accelerations can be calculated directly using the convolution,
\begin{equation}
f_i(\bx)=\int\rho(\bx')w_i(\bx-\bx')d^3x',
\end{equation}
where the anisotropic force kernels are given by
\begin{equation}
w_i(\bx)=-G\frac{x_i}{r^3}.
\end{equation}
The force method exactly reproduces the pair-wise inverse-square law on the grid, but comes at the cost of two extra FFTs.  The potential method is computationally less expensive, but the finite differencing degrades the pair-wise force resolution by a few grid cells.  While the pair-wise force is not exact at small separations, the net force on any given cell is in general still highly accurate.  In practice, the relatively small accuracy trade-off of the potential method is preferred over the factor of 2 increase in computational work and time of the force method.

In the operator-splitting scheme, the forward and reverse gravity operations are consecutive and can be considered as one operation.  For the gravity operation, we use the density field $\rho^{t+\Dt}$ to determine the gravitational potential $\phi^{t+\Dt}$, which is then finite differenced to give the acceleration field $\ba^{t+\Dt}$.  We impose the restriction that the time step must satisfy
\begin{equation}
\Dt<\frac{1}{\sqrt{{\rm max}(a_n)}},
\end{equation}
In practice, the gravitational time step constraint rarely enforces the limiting time step.  It tends to come into effect only when the fluid is initially at rest or moving relatively slowly so that the Euler time step is larger than the gravitational time step.

\section{Hydrodynamic Tests}
\label{sec:hydrotests}

In this section we apply the one-dimensional and three-dimensional versions of the {\it MACH} code to hydrodynamic tests like the Sod shock tube test and the Sedov-Taylor blast-wave test.  The accuracy of the relaxing TVD scheme has been previously tested extensively in \citet{pen98} and \citet{tp03}.  Here we are interested in demonstrating the strength of the {\it MACH} code to capture shocks in the presence of high Mach numbers.  We conduct two variants of each test.  The standard version is a static one, where the initial ambient medium is at rest with respect to the simulation box.  This version is often presented in the literature to showcase the accuracy of hydrodynamic schemes, but this idealized case minimizes the Mach number.  In practice, regions experiencing shocks are often dynamical and not at rest with respect to the simulation box.  In the modified version, the medium has an initial velocity boost and is rapidly moving through the periodic box.  The boosted test mimics the situation where the shock is embedded in a bulk flow and both preshock and postshock Mach numbers are high.

\subsection{Sod Shock Tube Test}

\begin{figure*}[t]
\epsscale{1.5}
\plotone{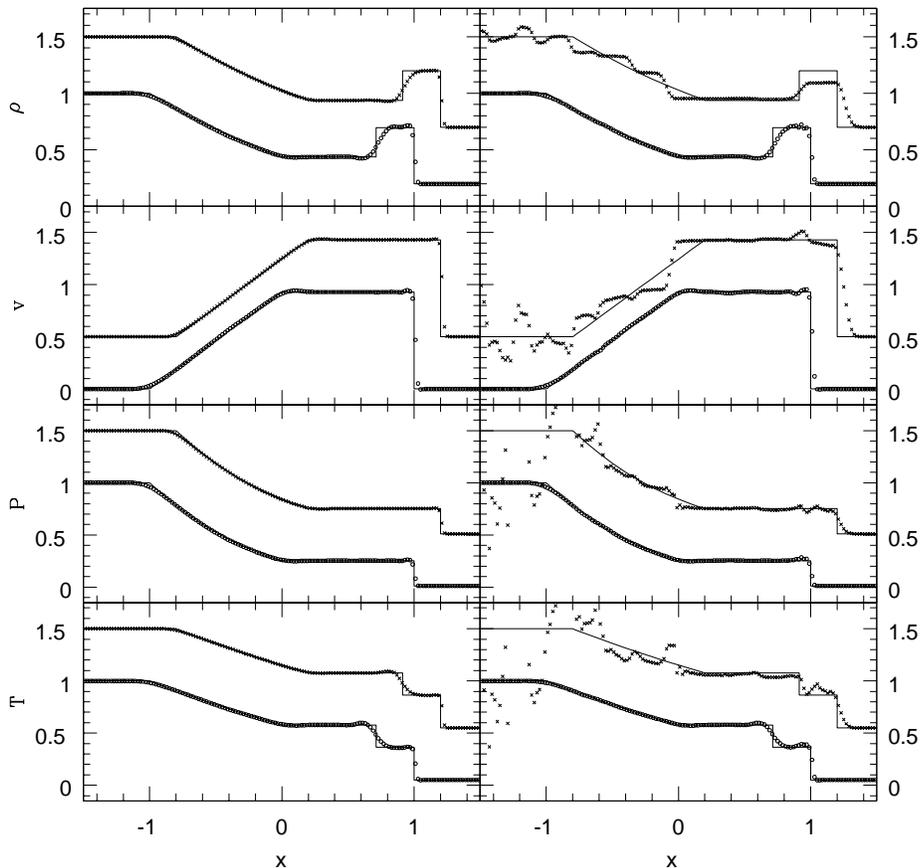}
\caption{Shock tube test results with the moving frame implementation turned on (open cicles) and off (crosses).  The standard test results are shown in the left panels and the boosted test results are found in the right panels.  In the boosted test, both preshock and postshock Mach numbers are high.}
\label{fig:shocktube}
\end{figure*}

The Sod shock tube is a special case of the Riemann problem, where a jump discontinuity in the initial conditions leads to the development of shocks.  Consider a one-dimensional tube containing two regions of fluid, initially at rest and separated by a membrane.  We label the initial state to the right of the membrane with the subscript 1 and the initial state to the left with the subscript 5.  We consider the case where $\rho_5>\rho_1$, $P_5>P_1$, and $v_5=v_1=0$.  At an initial time $t=0$, the membrane is instantaneously removed and the pressure imbalance results in a contact discontinuity and shock wave propagating to the right towards the low-pressure region and an expansion or rarefraction fan moving to the left towards the high-pressure region.  The shock, contact, and expansion each separate regions of steady flow that we label with the subscripts 2, 3, and 4 respectively.

The exact solution to the Riemann problem for the Euler equations uses the shock jump conditions and   self-similar arguments.  A thorough derivation is presented in \citet{laney98} and here we summarize some useful relations.  A contact discontinuity occurs when the velocity and pressure are continuous but other fluid properties are not.  Therefore, the velocity and pressure on both sides of the contact are equal and constant:
\begin{align}
v_2&=v_3,\\
P_2&=P_3.
\end{align}
The postshock pressure $P_2$ can be determined from the implicit equation
\begin{multline}
\frac{P_1}{P_5}=\frac{P_1}{P_2}\bigg[1-\frac{c_1}{c_5}\left(\frac{\gamma-1}{2\gamma}\right)\left(\frac{P_2}{P_1}-1\right)\\
\times\sqrt{\frac{2\gamma}{(\gamma+1)(P_2/P_1)+(\gamma-1)}}\bigg]^{2\gamma/(\gamma-1)},
\end{multline}
%\begin{equation}
%\frac{P_1}{P_5}=\frac{P_1}{P_2}\bigg[1-\frac{c_1}{c_5}\left(\frac{\gamma-1}{2\gamma}\right)\left(\frac{P_2}{P_1}-1\right)\sqrt{\frac{2\gamma}{(\gamma+1)(P_2/P_1)+(\gamma-1)}}\bigg]^{2\gamma/(\gamma-1)},
%\end{equation}
where $c_n=\sqrt{\gamma P_n/\rho_n}$ are the sound speeds.  The postshock velocity $v_2$ is then given by
\begin{equation}
v_2=\frac{c_1}{\gamma}\left(\frac{P_2}{P_1}-1\right)\sqrt{\frac{2\gamma}{(\gamma+1)(P_2/P_1)+(\gamma-1)}}.
\end{equation}
From the Rankine-Hugoniot shock jump relations, we obtain the shock velocity
\begin{equation}
v_s=\frac{c_1^2}{\gamma v_2}\left(\frac{P_2}{P_1}-1\right).
\end{equation}
The density jumps across the contact discontinuity and we have
\begin{align}
\rho_2&=\rho_1\frac{v_s}{v_s-v_2},\\
\rho_3&=\rho_5\left(\frac{P_3}{P_5}\right)^{1/\gamma}.
\end{align}
After time $t$, the contact discontinuity has propagated a distance of $x_c=v_3 t$ and the shock front is located at distance $x_s=v_s t$, relative to the initial membrane.  Region 4 is a rarefraction fan that decreases the density and pressure as it expands leftward towards the high pressure region.

The one-dimensional {\it MACH} alorithm is applied to the shock tube test with initial conditions: $\rho_5=1$, $P_5=1$, $\rho_1=0.2$, $P_1=0.01$.  The initial conditions are identical to \citet{shapiro95} and \citet{pen98}.  The pressure ratio $P_5/P_1=100$ is 10 times larger than that in \citet{sod78}, allowing a rigorous test of the code's ability to capture strong shocks at high Mach numbers.  In the static test, the fluid is initially at rest and $v_5=v_1=0$.  We run the simulation until the shock front has propagated a distance of $x_s=50$ grid cells.  In the boosted test, the fluid has uniform initial motion with $v_5=v_1=100c_1$.  In the elapsed time that it takes the shock to propagate a distance of $x_s=50$ grid cells in its own frame, the entire region of interest will have moved through the box a total distance of more than 20 times $x_s$.  We conduct these tests with the moving frame technique turned on and off.  In the latter case, the code reduces to the Relaxing TVD code \citep{pen98,tp03}.

In Figure (\ref{fig:shocktube}) we compare the results of the static and boosted tests.  In the left panels, the static test results for {\it MACH} code (open circles) and Relaxing TVD code (crosses) are plotted against the exact solution (solid lines).  The Relaxing TVD results have been shifted both horizontally and vertically for clarity.  The plots have been rescaled such that the initial discontinuity is placed at $x=0$ and the shock front at $x=1$.  The grid spacing corresponds to $\Dx=0.02$.  Both codes successfully capture the gas dynamics in this test.  The shock, contact, and expansion sharply mark the different regions discussed previously.  The shock front has been propagated accurately and resolved in roughly two grid cells, with no spurious oscillations.  The contact discontinuity has been degraded by diffusion, but this is inevitable when trying to advect such a front over approximately 35 grid cells.  Different flux limiters for the TVD scheme can give different results.  The superbee flux limiter has been found to be the least diffusive \citep[See][]{tp03}, but it is also the least stable.  From experience, the van Leer limiter provides a preferable combination of accuracy and stability.

The {\it MACH} and Relaxing TVD codes give very different results for the boosted test, as shown in the right panels.  The Relaxing TVD code suffers from the high Mach number problem in this test and its inability to accurately track the thermal energy is reflected in the relatively large jumps in the pressure and temperature curves.  In addition, the large bulk velocity imposes a small time step, which in turn means that a larger number of steps is required for the simulation to evolve to the chosen final time.  As a result, more diffusion is expected and both the shock and contact fronts are highly degraded.  The {\it MACH} code does not suffer from the high Mach number problem here.  Though the fluid is rapidly moving through the simulation box, the frame change in the {\it MACH} code is able to remove the bulk component, effectively reducing the problem so that the hydrodynamics can be solved like in the idealized static case.

\subsection{Sedov-Taylor Blast-Wave Test}

\begin{figure*}[t]
\epsscale{1.5}
\plotone{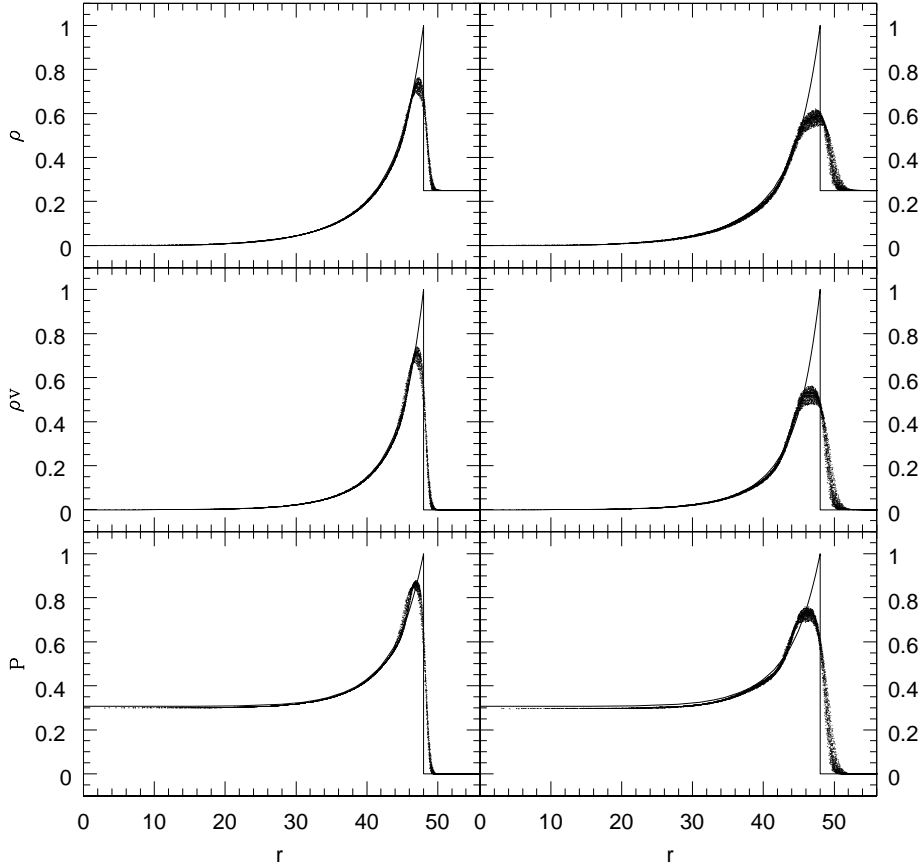}
\caption{Three-dimensional {\it MACH} algorithm applied to the standard (left panels) and boosted (right panels) Sedov-Taylor blast-wave tests.  In the boosted test, both preshock and postshock Mach numbers are high.}
\label{fig:sedovtaylor}
\end{figure*}

A rigorous and challenging test for the {\it MACH} algorithm is the three-dimensional Sedov-Taylor blast-wave test.  Consider an initially homogeneous medium with density $\rho_1$ and negligible pressure and instantaneously inject a point-like spike of thermal energy $E_0$ into the medium.  The explosion results in a spherical blast-wave that sweeps along material as it propagates outward.

The analytical Sedov solution uses the self-similar nature of the blast wave expansion and the full analytical solutions can be found in \citet{ll87}.  Here we collect some useful relations.  The propagation of the spherical shock front follows
\begin{equation}
r_{s}(t)=\xi_0\left(\frac{E_0 t^2}{\rho_1}\right)^{1/5}\ ,
\label{eqn:stre}
\end{equation}
where $\xi_0=1.15$ for an ideal gas with $\gamma=5/3$.  The velocity of the shock $v_{s}=\p r_{s}/\pt$ is given by
\begin{equation}
v_{s}(t)=\frac{2}{5}\frac{r_{s}(t)}{t}\ .
\end{equation}
The density $\rho_2$, velocity $v_2$, and pressure $P_2$ directly behind the shock front are:
\begin{align}
\rho_2&=\left(\frac{\gamma+1}{\gamma-1}\right)\rho_1\ ,\\[8pt]
v_2&=\left(\frac{2}{\gamma+1}\right)v_{sh}\ ,\\[8pt]
\ P_2&=\left(\frac{2}{\gamma+1}\right)\rho_1v_{sh}^2\ .
\end{align}
The immediate postshock gas density is constant in time, while the shocked gas velocity $v_2$ and pressure $P_2$ decrease as $t^{-3/5}$ and $t^{-6/5}$, respectively.

The three-dimensional {\it MACH} code is applied to two variants of the Sedov-Taylor blast-wave test.  We set up the simulation box with $128^3$ cells and constant initial density $\rho_1=1$.  At time $t=0$ we inject a spike of thermal energy $E_0=10^5$ into one cell at the centre of the box and let the simulation run until final time $t_f=35.59$ when the shock front has expanded to a radius of $r_{s}=48$ cells.  In the static test, the initial velocity of the medium is zero.  In the boosted test, all three components of the initial velocity is chosen to be equal to 100 times the sound speed $c_{48}=(\gamma P_2/\rho_2)^{1/2}$ of the immediate postshock gas at $r_{s}=48$ cells.  In the elapsed time, the centre of the explosion will have moved through the periodic box a total distance of more than 20 times the radius of the shock.  

In Figure (\ref{fig:sedovtaylor}) we compare the results of the standard and boosted tests.  In the standard test, the centre of the explosion is at rest with respect to the simulation box and the grid velocity is approximately zero.  In effect, the advection step contributes little to the fluid flow and the results are essentially the same compared to that produced by a standard Relaxing TVD code \citep[See][]{pen98,tp03}.  The shock is propagated accurately and resolved in roughly two grid cells.  There are no spurious oscillations and little anisotropic scatter is seen.  In the boosted test, the medium and the centre of the explosion are rapidly moving through the periodic box, but the frame change removes that bulk component, with the grid velocity being approximately equal to the initial velocity boost.  As a result, in the Euler step, the hydrodynamics is solved in a frame at rest with respect to the centre of the explosion, like that in the standard test.  The shock is correctly propagated despite the high Mach numbers.  There is more diffusion and scatter in this case compared to the standard one because of the additional advection step.  This is expected and inevitable when advecting over a large number of cells on an Eulerian grid.  In fact, the diffusion is minimized by the hybrid Lagrangian-Eulerian advection scheme implemented in the {\it MACH} code.  All previous Eulerian algorithms to date will fail this non-trivial boosted Sedov-Taylor test.

\section{Cosmological Hydrodynamics}
\label{sec:cosmo}

The {\it MACH} code has been adapted for cosmological applications in a Friedman-Robertson-Walker (FRW) universe.  In an expanding FRW background, we use comoving coordinates $x_c=x/a$, where the scale factor $a$ is governed by the Friedman equation
\begin{equation}
\left(\frac{da}{dt}\right)^2=a^2H_0^2(\Omega_m a^{-3}+\Omega_k a^{-2}+\Omega_\Lambda).
\end{equation}
The cosmological constants $H_0$, $\Omega_m$, $\Omega_k$, and $\Omega_\Lambda$ are present epoch values and following standard practice, we choose $a_0=1$ and obtain the relation
\begin{equation}
\Omega_m+\Omega_k+\Omega_\Lambda=1.
\end{equation}
In a cosmologically flat background, the curvature term $\Omega_k$ vanishes and we have $\Omega_{total}\equiv\Omega_m+\Omega_\Lambda=1$.

The standard cosmological hydrodynamic equations in comoving coordinates do not preserve the time-invariant conservation form of the Euler fluid equations.  However, with the appropriate spatial and time coordinate transformations, the conservation form can be maintained.  Following \citet{gnedin95} we introduce a new time variable
\begin{equation}
d\tau=\frac{dt}{a^2},
\end{equation}
which preserves the structure of the expanded Euler equations, but with $d\tau$ and $x_c$ in place of $dt$ and $x$, respectively.  The time-dependence of the cosmological expansion is included in the gravitational source term, where the Newtonian gravitational constant $G$ becomes $aG$.  This change is reflected in Poisson's equation and in the grid velocity equation.

The relationship between the scalefactor $a$, proper time $t$, and new time $\tau$ can be determined by considering the alternative Friedman equation
\begin{equation}
\left(\frac{da}{d\tau}\right)^2=a^3(\Omega_m H_0^2)\left[1+a\frac{1-\Omega_m-\Omega_\Lambda}{\Omega_m}+a^3\frac{\Omega_\Lambda}{\Omega_m}\right].
\label{eqn:friedman}
\end{equation}
In an Einstein-de Sitter universe where $\Omega_m=1$ and $\Omega_\Lambda=0$, we have the analytical solution
\begin{align}
a(t)&=\left(\frac{3H_0}{2}\right)^{2/3}t^{2/3},\\[8pt]
a(\tau)&=\left(\frac{4}{H_0^2}\right)(-\tau)^{-2},\\[8pt]
t(\tau)&=\left(\frac{16}{3H_0^4}\right)\tau^{-3}
\end{align}
where $-\infty<\tau<0$.  For a general cosmology with non-zero cosmological constant $\Omega_\Lambda$, there is no analytical solution.  In the {\it MACH} code, we integrate equation (\ref{eqn:friedman}) using a third-order Taylor expansion to obtain an accurate solution for $a(\tau)$.  The cosmological expansion also imposes a constraint on the numerical time step. We require that 
\begin{equation}
\frac{\Delta a}{a}<0.02,
\end{equation}
for every time step.  At high redshifts where the gas velocity and acceleration are relatively small, the expansion time step is generally smaller than both the Euler and gravitational time steps.

Under the spatial and time coordinate transformations, all density terms are for comoving volumes $d^3x_c=a^{-3}d^3x$.  The mass density $\rho(\tau)$ is per comoving volume and related to the proper mass density by
\begin{equation}
\rho=a^3\rho_p.
\end{equation}
The peculiar velocity $\bvel(\tau)\equiv d\bx_c/d\tau$ used in the code is related to the proper peculiar velocity $\bvel_p$ by the expression,
\begin{equation}
\bvel=a\bvel_p
\end{equation}
Correspondingly, the total energy density $e(\tau)$, pressure $P(\tau)$, and gravitational potential $\phi(\tau)$ are
\begin{align}
P=a^5 P_p,\\
e=a^5 e_p,\\
\phi=a^5 \phi_p,
\end{align}
where $P_p$, $e_p$, and $\phi_p$ are the proper pressure, proper peculiar total energy density, and proper peculiar potential.  One factor of $a^2$ comes the velocity transformation while the other factor $a^3$ comes from converting between comoving and proper volumes.

We now write down the unit conversions between grid values and physical values.  The unit conversions are chosen so that the grid quantities are close to unity.  We choose the grid spacing to be $\Dx=1$ and this sets the length unit,
\begin{equation}
{\cal L}=a\frac{L}{N},
\label{eqn:lengthunit}
\end{equation}
where $L$ is the physical length of the simulation box and $N$ is the number of grid cells per side length.
The mean comoving grid matter density is normalized to unity and this fixes the density unit, 
\begin{equation}
{\cal D}=a^{-3}\frac{3\Omega_mH_0^2}{8\pi G}.
\label{eqn:densityunit}
\end{equation}
The length and density unit together restrict the mass unit to be ${\cal M}={\cal D}{\cal L}^3$.  The time unit is uniquely determined once the grid value of the gravitational constant $G$ is set.  We choose $G=1/(6\pi)$ and obtain the time unit
\begin{equation}
{\cal T}=\frac{2a^2}{3}\frac{1}{\sqrt{\Omega_m H_0^2}}.
\label{eqn:timeunit}
\end{equation}
Note that this converts the grid time $\tau$ to physical time $t$.  The three fundamental  length, mass, and time units can be used to determine all other unit conversions.  The velocity unit is given by ${\cal V}={\cal L}/{\cal T}$ and the energy unit is ${\cal E}={\cal M}({\cal L}/{\cal T})^2$.

\citet{pen98} calls the new time variable $\tau$ a `Newtonian' time because with this substitution, Newton's laws apply directly.  Objects move in straight trajectories unless acted upon by an external force.  The scaling relations for adiabatic expansion are automatically preserved.  Consider a completely homogeneous universe with some initial density, velocity, and temperature.  The velocity $\bvel(\tau)$ will remain constant and the proper peculiar velocity will automatically scale as $\bvel_p\propto a^{-1}$.  Similarly, for constant $\rho(\tau)$ and $P(\tau)$, the temperature will scale as $T\propto a^{-3(\gamma-1)}$.

\subsection{Particle-Mesh N-body}

The evolution of the collisionless dark matter is tracked using the standard approach of the particle-mesh (PM) N-body scheme \citep[See][]{he88,efstathiou85}.  The collisionless particles are advected by solving Newton's equations of motion in an FRW universe.  For each particle, we store its comoving position $\bx(\tau)$ and velocity $\bvel(\tau)$ and these adherent quantities can be updated if the acceleration $\ba(\tau)$ is known.  The gravitational force field is computed using the total matter density field, with contributions from both the baryonic gas and dark matter particles.  Mass assignment of particles onto the grid is accomplished using the `cloud-in-cell' (CIC) interpolation scheme \citep{he88}.

We have constructed an explicit, second-order integration technique based on the Runge-Kutta scheme.  The gas and dark matter must be synchronized when the total density field is constructed.  Recall that in the operator-splitting scheme, the gravity operation is carried out inbetween the forward and reverse sections of the hydro sweep and the fluid has been been advanced from $u^\tau$ to $u^{\tau+\Dtau}$.  We first synchronize the gas and dark matter by advancing the particles,
\begin{equation}
\bx^{\tau+\Dtau}=\bx^t+\bvel^t\Dtau,
\end{equation} 
where this half-step is similar to the half-step in a standard Runge-Kutta scheme.  The particles are mapped to the grid using the CIC mass assignment scheme and the acceleration field is computed following the procedure outlined earlier in the section on gravitation ($\S$\ref{sec:gravity}).  The acceleration $\ba^{\tau+\Dtau}$ on each particle is interpolated from the grid using the CIC scheme.  Using the same CIC scheme to do mass assignment and force interpolation guarantees that no fictitious self-force is imposed on the particles.  The full-step in the Runge-Kutta scheme updates the position and velocity for each particle as follows:
\begin{align}
\label{eqn:xp}
\bx^{\tau+2\Dtau}&=\bx^\tau+\bvel^\tau(2\Dtau)+\ba^{\tau+\Dtau}\frac{(2\Dtau)^2}{2},\\[8pt]
\label{eqn:vp}
\bvel^{\tau+2\Dtau}&=\bvel^\tau+\ba^{\tau+\Dtau}(2\Dtau).
\end{align}
Note that $\ba^{\tau+\Dtau}$ can be expanded as $\ba^\tau+(d\ba/d\tau)\Dtau$ and with this substitution,  equations (\ref{eqn:xp}) and (\ref{eqn:vp}) will be identical to a Taylor expansion up to second-order.

This Runge-Kutta integration scheme for advancing the particles has several advantages.  It provides a way to couple and synchronize the gas and dark matter during the gravity step.  The time step $\Dtau$ can be adjusted every double sweep, just like in the operator-splitting scheme for hydrodynamics.  In addition, the position and velocity are synchronized at the start and end of every double time step and not staggered temporally like in the leap-frog scheme, for instance.

\subsection{Numerical Performance}

The cosmological {\it MACH} code is computationally fast, memory friendly, and efficiently parallelized.  The dimensional splitting technique for the hydrodynamics effectively decomposes the problem into a one-dimensional one and this has several advantages.  First, this allows us to write highly optimized, one-dimensional algorithms that are cache efficient.  Second, parallelization using OpenMP directives for shared-memory machines becomes straightforward.  During a hydro sweep, the grid can be divided into one-dimensional columns and these independent columns can be efficiently distributed amongst the multiple processors.  Third, the code is memory friendly since additional variables for the TVD and Runge-Kutta schemes only need to be stored temporarily in small arrays.  

The PM N-body algorithm is also fully parallelized.  For the first-half step in the Runge-Kutta scheme, the half-positions can be calculated in parallel with each processor responsible for a fraction of particles.  No extra memory overhead is needed for storing the intermediate positions.  The half-positions $\bx^{\tau+\Dtau}$ can be stored in the same array as the initial positions $\bx^\tau$.  The fully updated positions can be calculated as
\begin{equation}
\bx^{\tau+2\Dtau}=\bx^{\tau+\Dtau}+\bvel^{\tau+2\Dtau}\Dtau,
\label{eqn:xp2}
\end{equation}
where this equivalent alternative to equation (\ref{eqn:xp}) does not require the initial positions and velocities.  The mass assignment and force interpolation steps can be done in parallel by using the standard procedure of constructing linked lists \citep[See][]{he88} for the particles.

For a $1024^3$ grid, the {\it MACH} code requires 20 GB to hold the hydro array $\bu$.  Since the grid velocity is smooth by construction, we only need to store it at half the resolution and 1.5 GB is needed for the array $\btv$.  It is expanded to full resolution for computation, but when working on one-dimensional columns of data, this requires relatively little memory overhead.  Normally, we run simulations where the ratio of particles to grid cells is 1:8.  For $512^3$ particles, we require 3 GB to store both the positions and velocities and another 0.5 GB for the linked list.  For the gravity step, the total mass density field and potential can be stored in the same array and this requires another 4 GB.  In total for a $1024^3$ grid and $512^3$ particles, the {\it MACH} code uses 29 GB of memory to hold the large arrays and a few GB to hold smaller temporary ones.

\subsection{One-Dimensional Zeldovich Pancake Test}

\begin{figure*}[t]
\epsscale{1.5}
\plotone{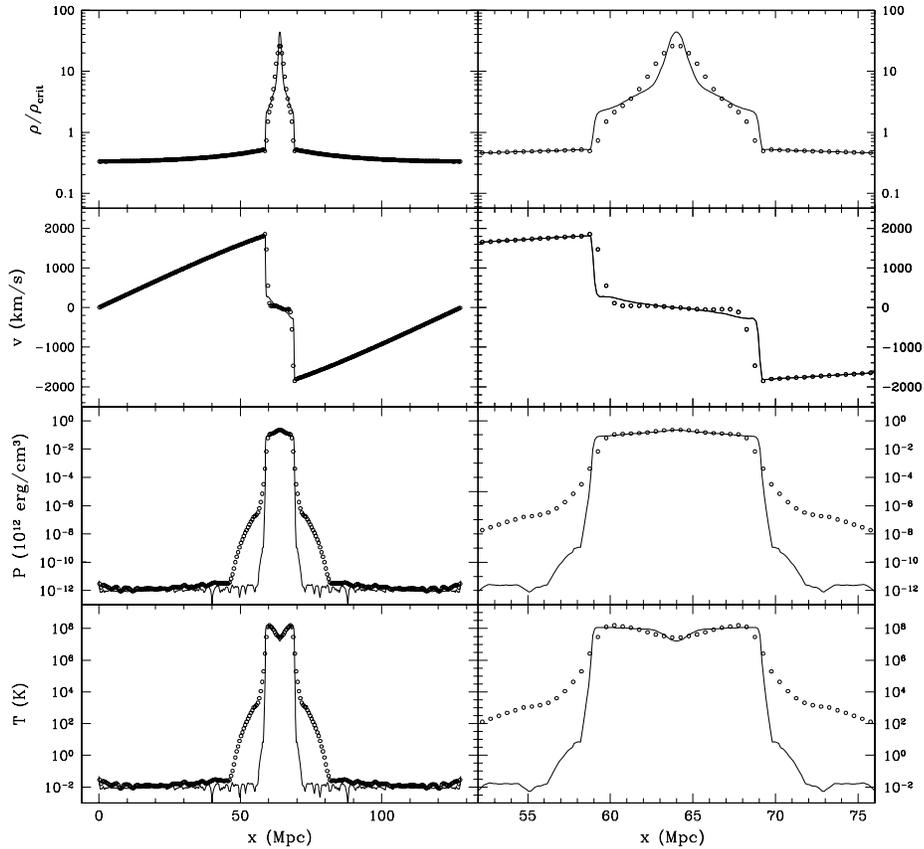}
\caption{Cosmological pancake test results at redshift $z=0$.  A lower resolution pancake from 256 cells (open circles) is plotted against a high resolution pancake with 1024 cells (solid line).}
\label{fig:pancake}
\end{figure*}

A standard cosmological test is the Zeldovich pancake problem \citep{zeldovich70}.  In the one-dimensional plane-parallel case, a sinusoidal perturbation evolves linearly initially and eventually collapses to form a caustic or pancake.  This vigorous test involves both hydrodynamics and self-gravity and strong shocks are formed in the presence of high Mach numbers.

For a flat cosmology, the initial conditions and the evolution of the sinusoidal perturbation \citep{zeldovich70,anninos94} can be determined from the following equations:
\begin{align}
\label{eqn:xq}
x(q,z)&=q-\frac{1+z_c}{1+z}\frac{\sin(kq)}{k},\\[8pt]
\rho(x,z)&=\rho_0\left[1-\frac{1+z_c}{1+z}\cos(kq)\right]^{-1},\\[8pt]
v_p(x,z)&=-H_0\frac{1+z_c}{\sqrt{1+z}}\frac{\sin(kq)}{k},
\end{align}
The solutions hold in both the linearly and moderately nonlinear regime prior to the redshift $z_c$ where the gravitational collapse forms a pancake.  For a given Eulerian comoving coordinate $x$, the corresponding Lagrangian coordinate $q$ can be calculated by inverting equation (\ref{eqn:xq}).  The comoving wavenumber $k=2\pi/\lambda$ sets the wavelength of the perturbation. The density $\rho$ is a comoving density and $\rho_0$ is the comoving average density.  The proper velocity $v_p$ has units of km/s.  For an ideal adiabatic gas, the thermodynamic solution is given by:
\begin{align}
T(x,z)&=T_i\left[\left(\frac{1+z}{1+z_i}\right)^3\frac{\rho(x,z)}{\rho_0}\right]^{\gamma-1}\\[8pt]
P(x,z)&=\frac{k_B}{\mu m_H}[(1+z)^3\rho(x,z)]T(x,z).
\end{align}
The temperature $T$ is a monotonic function of the density and $T_i$ is the initial average temperature at the initial redshift $z_i$.  The proper pressure $P$ is given by the equation of state for an ideal gas with mean molecular mass $\mu$.

We follow \citet{bryan95} and use the following parameters:  $z_i=100$, $z_c=1$, $\Omega_m=1$, $h=0.5$, $\lambda=64h^{-1}$ Mpc, $\rho_0=\rho_{crit}$, and $T_i=100$ K in a purely baryonic universe.  Initially, the maximum Mach number is $\sim 200$ and this makes the test challenging for an Eulerian code.  In Figure (\ref{fig:pancake}) we plot the nonlinear results at $z=0$.  A lower resolution pancake from 256 cells (open circles) is plotted against a high resolution pancake with 1024 cells (solid line).  For both runs, we used a Gaussian smoothing radius of 8 grid cells to compute the smoothed grid velocity. The formation of the pancake is due to two cold bulk flows colliding and collapsing to form a caustic.  The kinematics of the collapse is very well captured.  In the lower resolution run, the pancake is underresolved but the amount of diffusion is typical of that found by other second-order accurate TVD codes with fixed grids \citep[See][]{ryu93,pen98}.  In general for tests involving the formation of caustics, Eulerian schemes are more diffusive than Lagrangian schemes, which benefit from having high dynamic range in scale and density.  

The more difficult part of the pancake test is the thermodynamic evolution.   The initially cold gas is shock heated to very high temperatures near the caustic, while away from the shock the cold gas is expected to cool adiabatically as the universe expands.  Numerically, it is difficult to capture the large range in temperature, over 10 orders of magnitude for this particular test.  In addition, conventional Eulerian codes suffer from a large amount of spurious heating due to the high Mach numbers involved.    The moving frame approach allows us to remove the bulk flow and reduce the Mach numbers.  Away from the shock, the grid velocity approximates the bulk flow very well with $\tv\approx v$.  As a result, the Mach numbers are now small because the local velocity $\Dv$ is comparable to the low sound speed of the cold gas.  In Figure (\ref{fig:pancake}), we see no significant heating of the cold gas away from the shock.  The spurious heating has been minimized by adding the gravitationally induced accelerations to the grid velocity.  In the {\it PPM} code \citep{bryan95} there is an artificial minimum temperature floor of 1 K, but we impose no such restriction and the temperature floor in our simulations is consistent with the expected adiabatic cooling.  In the collapsing region, the grid velocity turns over and becomes zero at the center of the pancake.  In the postshocked region, the local velocity is large but the gas is very hot and therefore the Mach numbers are low.  The postshocked gas pressure and temperature are highly resolved.  Immediately outside of the expanding shock front, some heating is seen in the lower resolution run but not in the high resolution pancake.  In this region, the grid velocity turns over and the Mach numbers become high.

\subsection{Self-Similar Test}

\begin{figure*}[t]
\epsscale{1.5}
\plotone{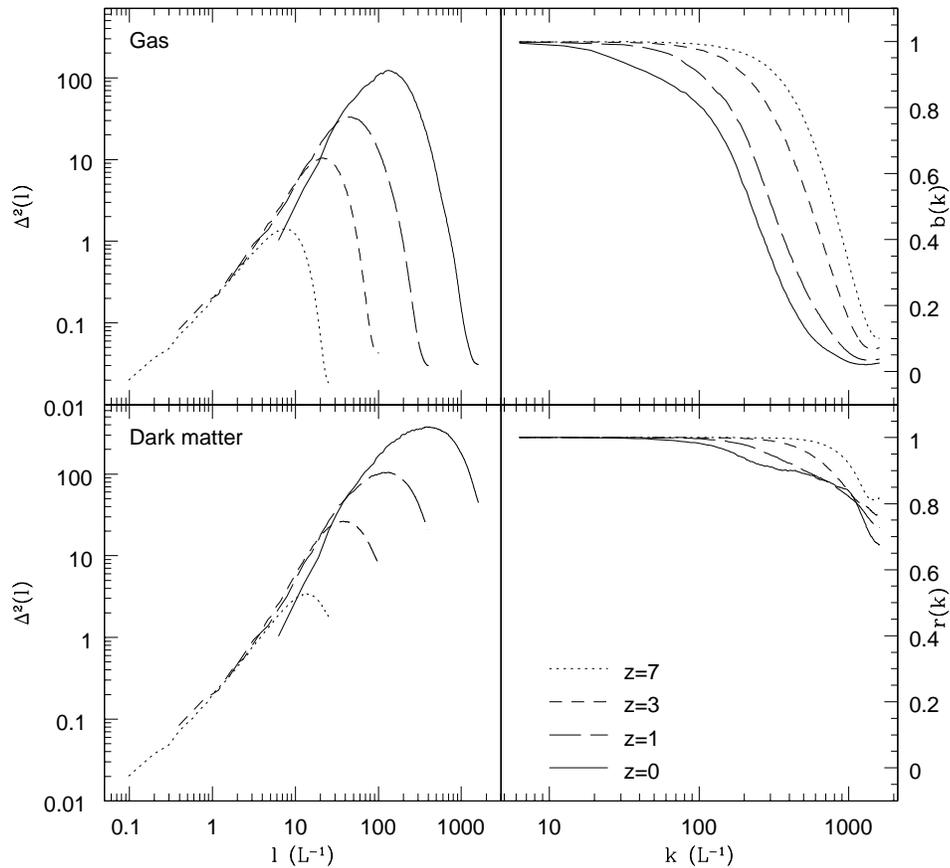}
\caption{Cosmological self-similar test for $n=-2$ scale-free initial conditions.  The scaled, dimensionless gas and dark matter power spectra are plotted in the left two panels while the bias and cross-correlation coefficient are plotted in the right panels.  The {\it MACH} code is run with $512^3$ grid cells and 1:1 gas to dark matter ratio.}
\label{fig:selfsimilar}
\end{figure*}

\begin{figure*}[t]
\epsscale{1.5}
\plotone{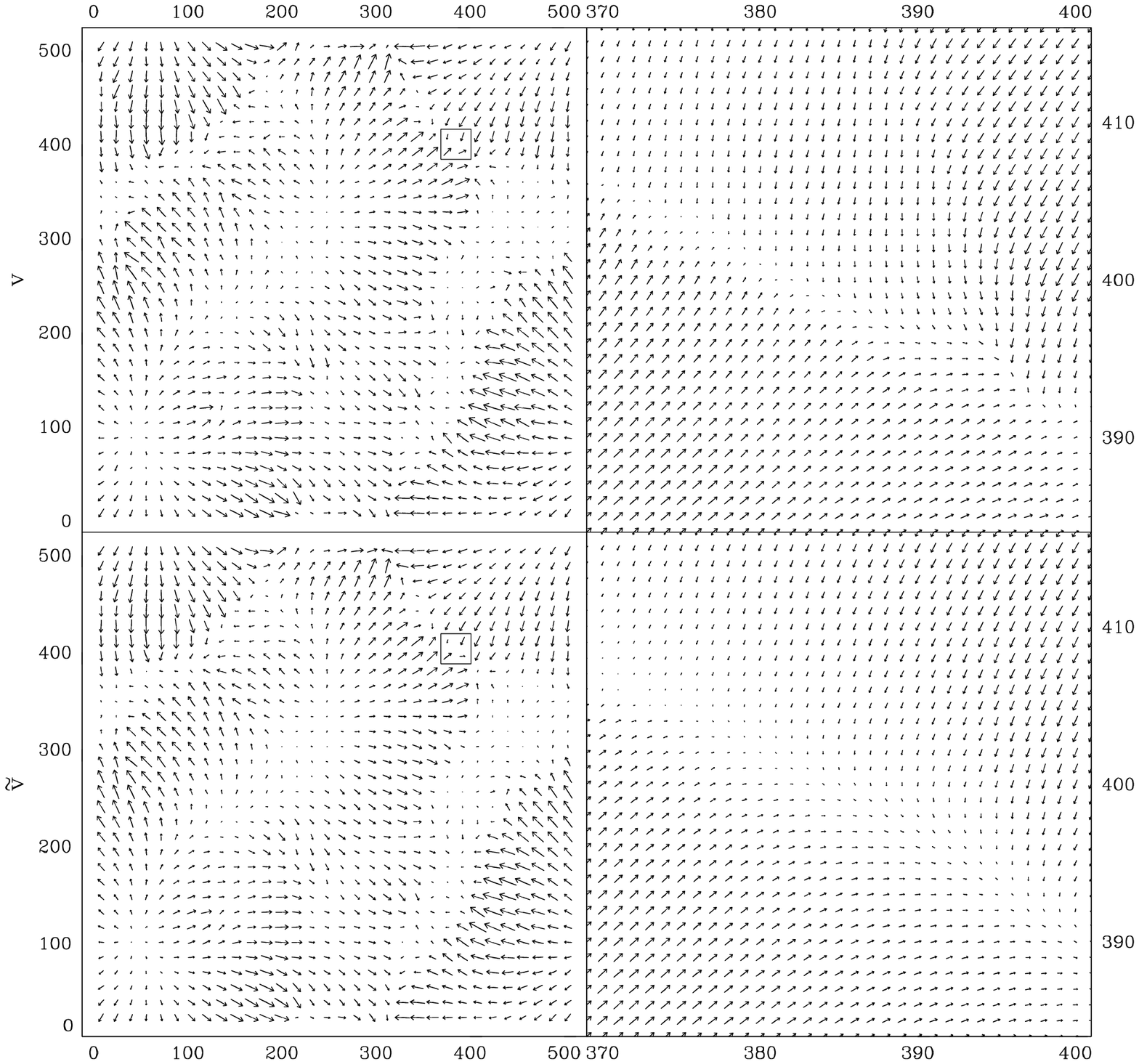}
\caption{Vector field diagram for slice at z=3.  The left panels compare the velocity field (top) and the smoothed grid velocity field (bottom) on the entire $512\times512$ grid.  The subsampled $32\times32$ vector field shows the large-scale bulk flow.  The full resolution right panels zoom in on a convergent region.}
\label{fig:velocity}
\end{figure*}

The self-similar evolution of scale-free cosmological initial conditions can be used to quantify the numerical limitations of hydrodynamic simulations.  In an Einstein-de Sitter universe where $\Omega_m=1$, scale-free initial conditions with power-law correlation function will evolve as
\begin{equation}
\xi(r)=\left(\frac{r_0}{r}\right)^\gamma\propto a^2.
\end{equation}
Hence, the correlation function obeys the self-similar transformation $\xi(r,t_1)\rightarrow\xi(s,t_2)$ where
\begin{equation}
\frac{s}{r}=\left[\frac{a(t_2)}{a(t_1)}\right]^{-2/\gamma}.
\end{equation}
In Fourier space, the power spectrum of the density modes will evolve as
\begin{equation}
P(k)\propto k^n\propto a^2,
\end{equation}
where $n=\gamma-3$.  In this case we have the transformation $P(k,t_1)\rightarrow P(l,t_2)$ where
\begin{equation}
\frac{l}{k}=\left[\frac{a(t_2)}{a(t_1)}\right]^{2/n}.
\end{equation}
Numerical artifacts arising from the finite grid resolution and the finite simulation volume will cause the evolution to deviate from the self-similar scaling.

We run a self-similar cosmological test with initial index $n=-2$ or equivalently $\gamma=1$.  We choose the normalization such that the linearly extrapolated correlation function at redshift $z=0$ has a correlation length $r_0$ equal to $1/4$ of the box length $L$.  The initial conditions are generated by first sampling the power spectrum to obtain $\delta(\bk)$, which is then inverse Fourier transformed to give the matter overdensity field $\delta(\bx)$ on the grid.  The initial particle displacement and velocity are found using the Zeldovich approximation \citep{zeldovich70}, as described by \citet{efstathiou85}.  The {\it MACH} code is run with $512^3$ grid cells and $512^3$ dark matter particles, starting from an initial redshift of $z=200$.  We choose the gas to matter ratio $\Omega_b/\Omega_m$ to be equal to the canonical ratio $1/6=0.05/0.30$ to approximate the current WMAP measurement \citep{spergel03}.

In Figure (\ref{fig:selfsimilar}) we plot the scaled gas and dark matter power spectra.  The dimensionless power spectrum is defined as
\begin{equation}
\Delta^2(k)=4\pi k^3 P(k),
\end{equation}
and $\Delta^2\sim1$ when the overdensity reaches the mildy nonlinear value $\delta\sim1$.  The self-similar scaling holds for a wide range of length and time scales.  The finite volume has no significant effect on the clustering at redshift $z=1$ where the correlation length is $r_0=L/16$.  At redshift $z=0$ where the correlation length is expected to be $r_0=L/4$, the clustering of large-scale structure is severely affected by finite volume effects and the self-similar scaling breaks down.

On large scales the gas is expected to be highly correlated with the dark matter and the evolution is driven primarily by gravity.  On small scales, gas pressure can be quite large due to gravitational collapse and shock heating and therefore we expect the gas to deviate from the dark matter.  We define the linear bias parameter as
\begin{equation}
b(k)=\sqrt{\frac{P_{gas}(k)}{P_{cdm}(k)}},
\end{equation}
and parametrize the correlation between the gas and the dark matter using the cross-correlation coefficient
\begin{equation}
r(k)=\frac{P_{gm}(k)}{\sqrt{P_{gas}(k)P_{cdm}(k)}},
\end{equation}
where $P_{gm}(k)\equiv\langle\delta_{gas}(k)\delta_{cdm}(k)\rangle$ is the cross-spectrum.  The cross-correlation coefficient or stochasticity parameter can have values $-1\leq r\leq 1$.  Figure (\ref{fig:selfsimilar}) shows that the correlation is perfect at the largest scales and weakens slightly at smaller scales.  The bias diminishes with decreasing scale since gas pressure counteracts gravitational infall and thereby reduces gas power.  The bias and stochasticity curves should be identical at all times if self-similarity holds.  However, self-similar scaling is broken at small scales because of finite grid resolution.  In addition, gravity introduces a length scale since the CIC mass assignment scheme and the finite-differencing force scheme together soften the force resolution.

An interesting exercise is to plot a vector field diagram comparing the velocity field $\bvel$ and the grid velocity field $\btv$.  We take an $x-y$ planar slice and use the $x$ and $y$ components of the velocities to construct the vector field diagram shown in Figure (\ref{fig:velocity}).  For the grid velocity, the $x$ component is obtained by smoothing in the $x$ direction and likewise for the $y$ component.  In Figure (\ref{fig:velocity}) the left panels show the vector fields on the entire $512\times512$ grid, but with subsampling so that only $32\times32$ vectors are shown.  The subsampling has the effect of displaying only the large-scale bulk flow.  Overall the grid velocity approximates the bulk flow very well.  In the right panels we zoom in a convergent region.  At full resolution on smaller scales, the grid velocity appears very smooth, a requirement for the algorithm to work accurately.

\section{Future Work}

The cosmological {\it MACH} code is highly suited for simulating the evolution of the IGM where accurate thermodynamic evolution is needed for studies of the Lyman alpha forest, the Sunyaev-Zeldovich effect, and the X-ray background.  The code has been equipped with a radiative transfer algorithm written by Uros Seljak following the prescription of \citet{cen92} and \citet{theuns98}.

We are currently implementing the {\it MACH} algorithm to run out-of-core hydrodynamic simulations (Trac \& Pen 2003 in prep).  Out-of-core computation refers to the idea of using disk space as virtual memory and transferring data in and out of main memory at high I/O bandwidth.  We can run high-resolution cosmological simulations with up to $4000^3$ grid cells and $2000^3$ particles on a 32 processor Alpha server with a 3 terabyte SCSI disk array at the Canadian Institute for Theoretical Astrophysics (CITA).

\section{Conclusions}

We have presented a hybrid hydrodynamic algorithm that solves the Eulerian fluid conservation equations in an adaptive frame moving with the fluid where Mach numbers are minimized.  Using a velocity decomposition technique, we define a local velocity and local total energy density and store the bulk kinetic components in a smoothed background velocity field that is associated with the grid velocity.  In addition, gravitationally induced kinetic changes are added to the grid rather than to the local quantities, thereby minimizing the spurious heating problem plaguing cold gas flows.  Separately tracking local and bulk flow components allows thermodynamic variables to be accurately calculated in both subsonic and supersonic regions.  A main feature of the algorithm is the ability to resolve shocks and prevent spurious heating where both the preshock and postshock Mach numbers are high.  

The {\it MACH} algorithm has been subjected to a one-dimensional Sod shock tube test and a three-dimensional Sedov-Taylor blast-wave test.  We have also modify these standard tests to include an initial velocity boost.  The modified tests mimic the situation where the shock is embedded in a supersonic flow where both preshock and postshock Mach numbers are high.  The moving frame algorithm is able to accurately resolve the shocks without forming spurious oscillations in all tests conducted.  Previous Eulerian implementations will fail the non-trivial boosted tests where the postshock Mach numbers remain high.

For cosmological applications, we have combined the moving frame algorithm with a PM N-body scheme to simultaneously capture the hydrodynamic evolution of the baryonic gas and the gravitational evolution of the collisionless dark matter particles.  In the cosmological pancake test, the code successfully captures both the nonlinear kinematic and thermodynamic evolution.  Although the temperature ranges over 10 orders of magnitude for this particular test, the thermodynamic profiles are highly resolved.  Away from the shock, the adiabatic cooling of the cold gas is accurately simulated with no spurious heating seen.  The {\it MACH} code does not suffer from the high Mach number problem encountered in previous Eulerian hydrodynamic codes.  Tests conducted with cale-free cosmological initial conditions show that the self-similar scaling laws hold over a wide range of length and time scales.

\section{Acknowledgments}

We thank Dongsu Ryu for interesting discussions and the suggestion to use a temperature weighting scheme when smoothing the velocity field.  HT wishes to thank Uros Seljak and Gasper Tkacik for friendly hospitality during a visit to Princeton University where a radiative transfer algorithm was added to the {\it MACH} code.




\end{document}